\documentclass{iopart}
\usepackage{iopams}
\usepackage{amssymb,amstext}
\usepackage{graphicx}
\usepackage[sort&compress,comma,square]{natbib}

\def\kB{k_{\text{B}}}
\def\kT{\kB T}
\let\bs\boldsymbol

\begin{document}

\title[The Dynamics of Silica Melts under High Pressure]%
  {The Dynamics of Silica Melts under High Pressure:
  Mode-Coupling Theory Results}
\author{Th~Voigtmann and J~Horbach}
\address{Institut f\"ur Materialphysik im Weltraum,
  Deutsches Zentrum f\"ur Luft- und Raumfahrt (DLR), 51170 K\"oln, Germany}
\eads{\mailto{thomas.voigtmann@dlr.de}}

\begin{abstract}
The high-pressure dynamics of a computer-modeled silica melt is studied
in the framework of the mode-coupling theory of the glass transition (MCT)
using static-structure input from molecular-dynamics (MD) computer simulation.
The theory reproduces the experimentally known viscosity minimum (diffusivity
maximum) as a function of density or pressure and explains it in terms of
a corresponding minimum in its critical temperature. This minimum arises
from a gradual change in the equilibrium static structure which shifts from
being dominated by tetrahedral ordering to showing the cageing known from
high-density liquids. The theory is in qualitative agreement with computer
simulation results.
\end{abstract}

\section{Introduction}

The physical mechanisms behind vitrification are still widely debated.
Studies taking into account the high-pressure behaviour of glass-forming
liquids in addition to their response to temperature variation, are now
emerging as a valuable means to provide insight into the glass transition
phenomenon.
Based on a comparison of colloidal and molecular glass-transition
data over large pressure ranges, it has been proposed that pressures well
above $1\,\text{GPa}$ are needed to significantly change the properties
of many well-studied fragile glass formers
\cite{tvpoon}; only in this extreme pressure
regime one would have hope to resolve the long-standing debate whether
energetic or entropic interactions are the main cause for the dynamical
slowing down in the vicinity of the glass transition.

It has been pointed out \cite{Brazhkin.2006} that most organic glass formers
will cease to exist as such under these conditions, the molecules being
irreversibly transformed to polymerized modifications. Even some of the
existing measurements on organic glass formers,
touching on the $1\,\text{GPa}$ regime
\cite{Cook.1994}, might have to be reconsidered in light of this
finding.\footnote{Brazhkin~V~V, private communication}
This leaves two material classes as possible
candidates for furthering the understanding of pressure-induced vitrification
and, by this route, the physics of the glass transition itself:
metallic glasses on the one hand, and amorphous silica and relatives on the
other.

Silica and silicate melts are special in that they are both ubiquitous in
application and known to display `anomalous'
changes in thermodynamic and kinetic properties as pressure is increased.
This arises essentially because they exist as open tetrahedral-network-forming
structures under atmospheric conditions and can be classified as
`strong' liquids in Angell's sense.
A prominent feature in silica and a number of alkali silicates
is that the diffusivity of the $\text{Si}$ and
$\text{O}$ atoms first \emph{increases} with increasing pressure
\cite{Kushiro.1978,Rubie.1993,Poe.1997,Tinker.2003}, contrary
to what one expects from the excluded-volume picture of the glass transition,
where increased pressure, insofar as it leads to increasing density,
drives dynamic arrest. Only at pressures higher than about
$10$--$20\,\text{GPa}$, depending on the melt composition, do the
diffusion coefficients decrease for silica melts, leading to a maximum
in the diffusivity-versus-pressure plot, as predicted from
computer simulation \cite{Angell.1982,Barrat.1997,Shell.2002}.
Similarly, the viscosity first
decreases with increasing pressure, eventually showing a minimum.

The mode-coupling theory of the glass transition, MCT \cite{leshouches},
is commonly accepted to be a theory applicable to relatively high
temperatures in `fragile' liquids where it has been tested with great
success \cite{goetze99}, although MCT signatures such as two-step slow
relaxation have also been seen in `strong' liquids displaying nearly-Arrhenius
behaviour at low temperatures \cite{Sidebottom1993}. Earlier simulation and
combined simulation-and-theory studies \cite{sciortino01,Voigtmann2006}
point out by demonstrating qualitative agreement with microscopic MCT
calculations
for silica and sodium silicate mixtures, that the theory can in fact
yield more detailed predictions also for strong glass-forming liquids.
The existence of a MCT-$T_c$ also for silica implies a number of asymptotic
predictions, for example that the long-time `structural relaxation' in
the system is in fact independent on the details of microscopic motion;
this strong MCT prediction has recently been confirmed in a simulation
study comparing molecular-dynamics with stochastic Monte-Carlo dynamics
for a simulation model of silica \cite{berthier}.
Still, it remains
a crucial question towards understanding the theory's benefits and limitations,
how MCT deals with the differences between `strong' and `fragile' liquids.
Silica has been argued to undergo a transition from one to the
other upon pressurization \cite{Barrat.1997},
and so is an ideal candidate for these studies.

In this contribution, we present first MCT results for pressurized
silica melts. We demonstrate that the diffusivity maximum is reproduced
by the theory, using computer-simulated static structure factors as input.
The maximum of diffusivity corresponds to a minimum of the MCT critical
temperature $T_c$ as a function of density $\varrho$ or pressure $P$,
which in turn arises from the interplay of decreasing tetrahedral-network
effects and increasing contributions from nearest-neighbour cageing.

\section{Model Calculations}

Since MCT predicts drastic changes in the dynamics arising from
relatively minor changes in the average equilibrium structure, one needs
to ensure good-quality input for the latter. For complex liquids such as
silica, this can be delivered by computer simulation in conjunction with
a reliable model potential. Carr\'e \etal\ \cite{CHIK} have
recently developed a pair potential (called CHIK potential)
based on Car-Parrinello calculations
that reproduces the experimental equation of state reliably and hence improves
significantly over the Beest-Kramer-van~Santen (BKS) potential \cite{Beest1990}
widely used so far.
The CHIK potential was used in extensive molecular-dynamics (MD) computer
simulations described in detail elsewhere in this issue
\cite{Horbach.thisissue}. From these, the matrix of equilibrium partial
static structure factors $\bs S(q)=\langle n_\alpha(\vec q,t)^*
n_\beta(\vec q,0)\rangle$ was obtained, where $n_\alpha(\vec q,t)
=\sum_i\exp[i\vec q\cdot\vec r_{i,\alpha}(t)]$ are the Fourier-transformed
(wave vector $\vec q$) number-density fluctuations
of species $\alpha=\text{Si},\text{O}$, and the
sum runs over all $N_\alpha$ particles belonging to that type with
positions $\vec r_{i,\alpha}(t)$.
For use in the MCT equations, $\bs S(q)$ has been simulated for
the isotherms $T=2100\,\text{K}$, $2230\,\text{K}$, $2580\,\text{K}$,
$2750\,\text{K}$, $3250\,\text{K}$, and $3580\,\text{K}$ at various densities
from $2.3\,\text{g}/\text{cm}^3$ to $4.3\,\text{g}/\text{cm}^3$;
at other temperatures, $\bs S(q)$ was obtained by
linear interpolation between the above values unless otherwise noted.

MCT equations of motion for the resulting binary mixture are solved
numerically as outlined in Ref.~\cite{goetze03}. They yield the matrix of
dynamic partial number-density correlation functions $\bs\Phi(q,t)$
depending on wave number $q=|\vec q|$,
\begin{equation}\label{mct}
\eqalign{
  \bs J^{-1}(q)\partial_t^2\bs\Phi(q,t)
  +\bs S^{-1}(q)\bs\Phi(q,t)\\
  +\int_0^t\bs M(q,t-t')\partial_{t'}\bs\Phi(q,t')\,dt'=\bs0\,,
}
\end{equation}
where $J_{\alpha\beta}(q)=q^2\kT/m_\alpha\delta_{\alpha\beta}$ sets the
thermal velocities for the short-time dynamics
and $\bs\Phi(q,0)=\bs S(q)$.
$\bs M(q,t)$ is the memory function matrix of generalized fluctuating
forces,
which in the MCT approximation is written as
\begin{equation}
\eqalign{
  M_{\alpha\beta}(q,t)=\frac1{2q^2}\frac{n}{x_\alpha x_\beta}
  \int\frac{d^3k}{(2\pi)^3}
  \times\\ \times
  \sum_{\alpha'\beta'\alpha''\beta''}
  V_{\alpha\alpha'\alpha''}(\vec q,\vec k)V_{\beta\beta'\beta''}(\vec q,\vec k)
  \Phi_{\alpha'\beta'}(k,t)\Phi_{\alpha''\beta''}(p,t)
}
\end{equation}
with $p=|\vec q-\vec k|$. Here, $n$ is the number density, and
$x_\alpha$ are the number concentrations, $x_{\text{Si}}=1/3$ and
$x_{\text{O}}=2/3$ in our case. The vertices
$V_{\alpha\alpha'\alpha''}(\vec q,\vec k)
=(\vec q\vec k/q)c_{\alpha\alpha'}(k)\delta_{\alpha\alpha''}
+(\vec q\vec p/q)c_{\alpha\alpha''}(p)\delta_{\alpha\alpha'}
+qn x_\alpha c^{(3)}_{\alpha\alpha'\alpha''}(\vec q,\vec k)$
contain only equilibrium static correlations, viz.\ the matrix of
direct correlation functions $\bs c(q)$ defined by $\bs S(q)$ through the
Ornstein-Zernike relation
\cite{HansenMcDonald}. $c^{(3)}$ denotes the corresponding static triplet
correlation function \cite{triplet}, which we set to zero in
the following since no simulation data for it are available so far.
This neglect of triplet correlations has been studied in
detail for silica melts modeled through the BKS potential
\cite{sciortino01}, where $c^{(3)}$ was found to give noticeable contributions
different than in, for example, dense Lennard-Jones mixtures.
Without triplet contributions, the silica results under atmospheric
pressure were still qualitatively correct regarding the wave-vector
dependence of the correlation functions. We expect the same situation
for our model potential, with the quality of this additional approximation
to become better on increasing density.

From the resulting $\bs\Phi(q,t)$, an equation similar to Eq.~(\ref{mct})
allows to calculate the $\alpha$-species tagged-particle density correlations
$\phi_\alpha(q,t)=
\langle\exp[i\vec q\cdot(\vec r_\alpha(t)-\vec r_\alpha(0))]\rangle$, where
$\vec r_\alpha(t)$ marks a single tracer particle. The corresponding
MCT memory kernel reads $m^s_\alpha(q,t)=(1/q^2)\int d^3k/(2\pi)^3\,
\sum_{\alpha'\beta'}(\vec q\vec k/q)^2 c_{\alpha\alpha'}(k)
c_{\alpha\beta'}(k)\Phi_{\alpha'\beta'}(k,t)\phi^s_\alpha(p,t)$. In the
limit $q\to0$, the quantity $q^2m^s_\alpha(q,t)$ approaches a finite limit
which plays the role of the memory kernel for the corresponding
mean-squared displacement $\delta r^2_\alpha(t)$. The self-diffusion
coefficients $D_\alpha=\lim_{t\to\infty}\delta r^2_\alpha(t)/(6t)$ can
thus be determined as $D_\alpha^{-1}=\int_0^\infty dt \lim_{q\to0}
q^2m^s_\alpha(q,t)$.
For the numerical solution of Eq.~(\ref{mct}) and the corresponding
equations determining $\phi^s_\alpha(q,t)$ and $\delta r_\alpha^2(t)$,
we use a wave-vector grid with a cutoff of
$Q=24/\text{\AA}$ and grid spacing $\Delta q=0.1/\text{\AA}$.

MCT describes the slowing down of diffusivity, $D_\alpha\to0$,
connected to an increase in relaxation times for the $\bs\Phi(q,t)$,
as the coupling described by the $V_{\alpha\alpha'\alpha''}$ increases
smoothly through a variation of control parameters $(\varrho,T)$.
The divergence of relaxation times defines the MCT critical point
$T_c(\varrho)$. No such divergence is observed in experiment or simulation,
but the scaling laws connected to $T_c$ describing the asymptotic shape
of the correlation functions and a power-law variation in relaxation
times are \cite{goetze99}. Hence, the MCT critical
point provides a useful, well-defined concept to discuss the slow dynamics
of glass-forming systems.

\section{Results}

\begin{figure}
\begin{indented}\item[]
\includegraphics[width=\linewidth]{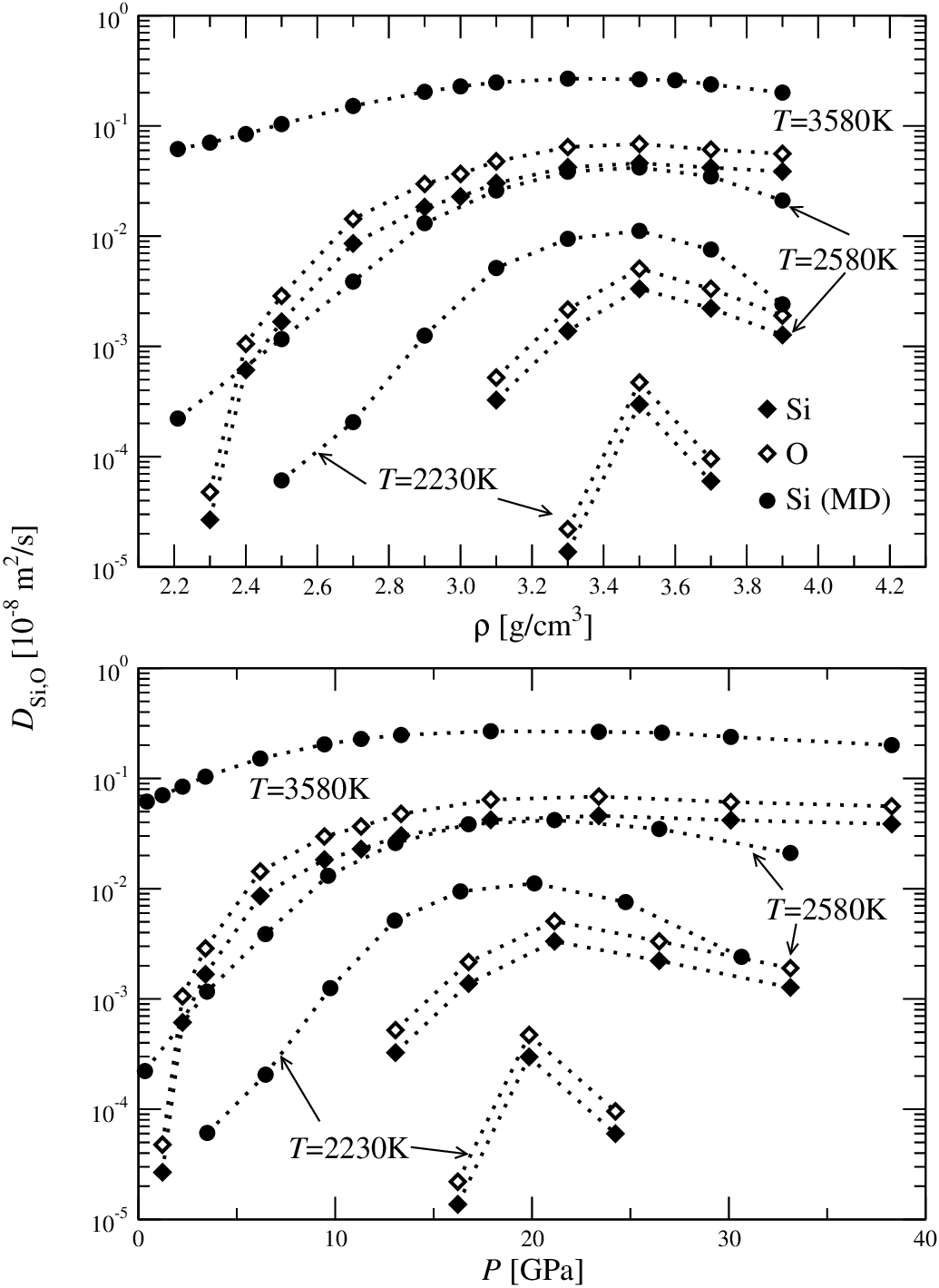}
\end{indented}
\caption{\label{fig:diff}%
  Self-diffusion coefficients of silicon, $D_{\text{Si}}$ (filled diamonds),
  and oxygen, $D_{\text{O}}$ (open diamonds), in model silica melts calculated
  by the mode-coupling theory of the glass transition along the indicated
  isotherms, as functions of density $\varrho$ (top) and as functions
  of pressure $P$ (bottom). Filled circles: $D_{\text{Si}}$ obtained from
  molecular-dynamics simulation.
}
\end{figure}

Figure~\ref{fig:diff} shows the self-diffusion coefficients $D_\alpha$
($\alpha=\text{Si},\text{O}$) calculated from MCT
along isotherms as diamond symbols. A maximum is found for all isotherms
considered, at a density of about $\varrho_0\approx 3.50\,\text{g}/\text{cm}^3$
roughly independent of temperature. This corresponds to a diffusivity maximum
at a pressure of about $P_0\approx 20\,\text{GPa}$, as the lower panel
of Fig.~\ref{fig:diff} shows. There, density values have been translated
to pressure by means of the computer-simulated equation of state
\cite{CHIK,Horbach.thisissue}. In both representations, the maximum is
more pronounced at lower temperatures, indicating that it is a feature
of slow glassy dynamics. At high temperatures, only the pronounced initial
increase of $D_\alpha$ with pressure remains clearly visible, while the region
around the maximum becomes rather broad.
Note that all the temperatures discussed here are
still high compared to the conventional glass transition $T_g$.
The maximum in diffusivity also corresponds to a minimum in viscosity,
or more generally a minimum in structural relaxation times.
For $\text{O}$ diffusion, we find the same behaviour as for $\text{Si}$
diffusion, while $D_{\text{O}}$ is slightly larger than $D_{\text{Si}}$;
the ratio $D_{\text{O}}/D_{\text{Si}}$ for $T=3580\,\text{K}$ drops from
roughly $1.78$ at $\varrho=2.3\,\text{g}/\text{cm}^3$ to about $1.45$
at $\varrho=4.2\,\text{g}/\text{cm}^3$; other isotherms give similar
behaviour. These ratios are slightly larger than what has been measured
in a silicate melts with several alkali-oxyde additions \cite{Tinker.2003}
at diffusivities of ${\mathcal O}(10^{-11}\,\text{m}^2/\text{s})$.
Note however that at lower diffusivities, the transport of $\text{Si}$
and $\text{O}$ will be governed by hopping processes with different
activation energies, leading to a much larger $D_{\text{O}}/D_{\text{Si}}$.

The diffusion coefficients from MCT are in qualitative agreement with
results from MD simulation using the CHIK potential.
This is demonstrated for $D_{\text{Si}}$,
where Fig.~\ref{fig:diff} reproduces some simulation data from
Ref.~\cite{Horbach.thisissue} as circle symbols. In particular, the
density and pressure of maximum diffusivity found in MCT correspond
well to the computer-simulation results. Already earlier simulation studies
based on the BKS potential for silica have reported a diffusivity maximum
at a density of
$\varrho_0^{\text{MD}}\approx3.5\,\text{g}/\text{cm}^3$
\cite{Barrat.1997,Shell.2002} indicating that this feature is robust
against slight changes in the potential. The absolute values of $D_{\text{Si}}$
disagree between MD and MCT. In particular at low temperatures and low
densities, additional relaxation processes not captured
in the theory render the divergence of the diffusion coefficient much
weaker than predicted by the theory, hence the disagreement is most pronounced
in this regime. Here, computer simulations show a temperature dependence
of the $D_\alpha$ that is well described by Arrhenius laws, which are not
reproduced in MCT.

\begin{figure}
\begin{indented}\item[]
\includegraphics[width=\linewidth]{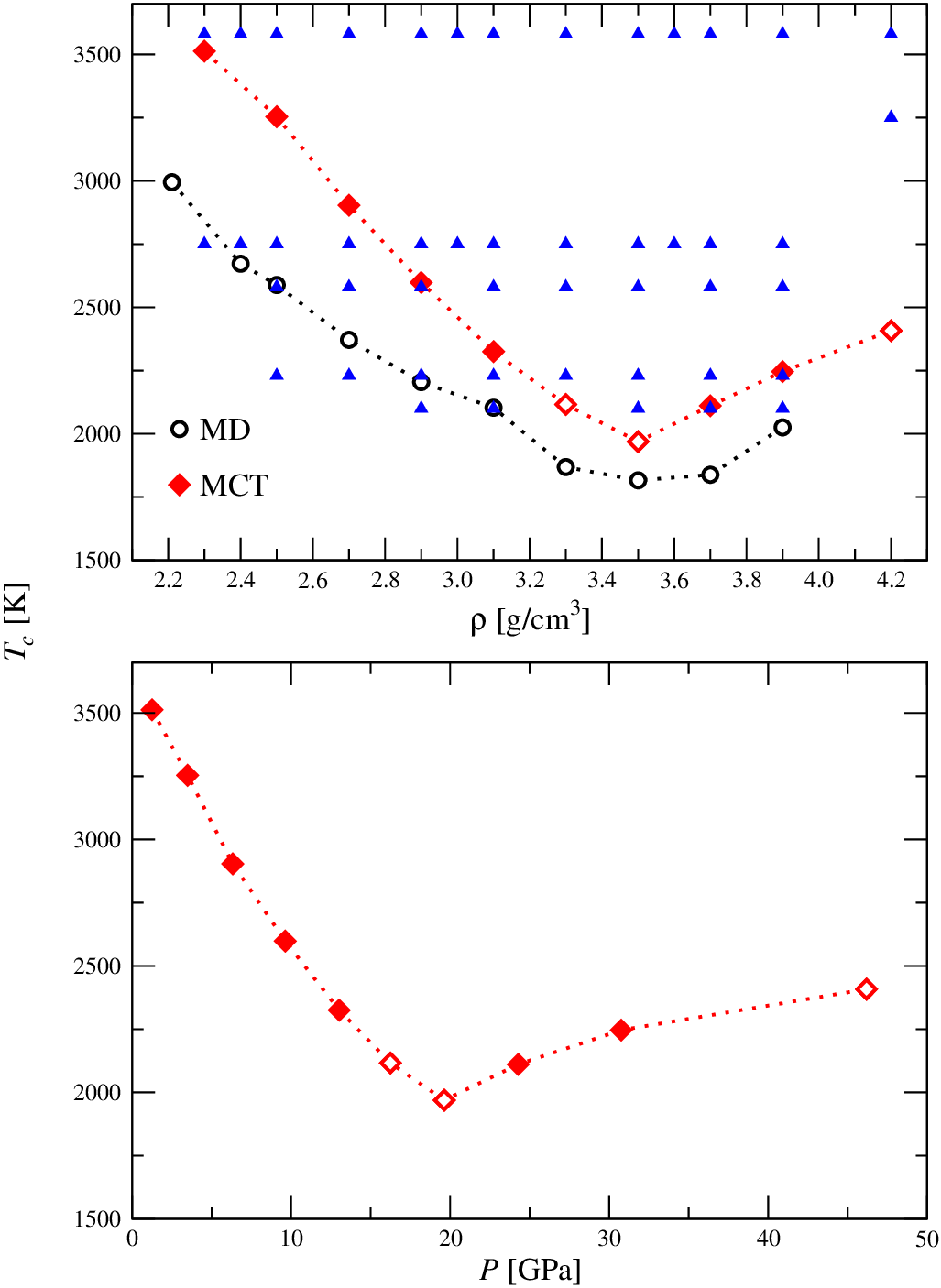}
\end{indented}
\caption{\label{fig:tc}%
  Mode-coupling-theory critical temperature $T_c$ for silica, calculated
  by MCT based on MD-simulated static structure factors with the
  CHIK potential (diamonds), and estimated from the simulation (circles).
  Open diamonds are calculated by extrapolation of MD structure factors
  $\bs S(q)$ available only for $T>T_c$;
  small triangles indicate state points for which $\bs S(q)$ was available
  from MD.
}
\end{figure}

The strong variation of $D_\alpha(\varrho)$ -- it increases by orders of
magnitude upon increasing pressure up to $P_0$ --
suggests an explanation in terms of a corresponding variation of
the critical temperature $T_c(\varrho)$ of mode-coupling theory.
The values calculated from a bifurcation analysis of the long-time limit
of Eq.~(\ref{mct})
are shown in Fig.~\ref{fig:tc}. Indeed, $T_c(\varrho)$ shows a pronounced
minimum around $\varrho_0$.
For most densities, it was possible in the simulation to
obtain equilibrated configurations also below this $T_c$, so that the
calculation
could be based on linear interpolation of a set of state points for
which $\bs S(q)$ was simulated, and
between which the structure factor changes were small. Only around
$\varrho_0$, the critical temperature is too low to equilibrate the system
within reasonable time scales in MD.
Additionally, at
$\varrho=4.2\,\text{g}/\text{cm}^3$ and temperatures around and below
$T=2750\,\text{K}$, the simulations showed crystallization, preventing
access to the liquid regime. Therefore, the $T_c$ values shown for
$\varrho=3.3\,\text{g}/\text{cm}^3$, $3.5\,\text{g}/\text{cm}^3$, and
$4.2\,\text{g}/\text{cm}^3$ are based on extrapolation of $\bs S(q)$ from
higher $T$ and have higher uncertainty than the rest of the
$T_c$ data; the state points at which `exact' MD input for $\bs S(q)$
was used are marked in Fig.~\ref{fig:tc} by diamond symbols.
Translating the $(\varrho,T_c)$ pairs to
$T_c(P)$ by the simulated equation of state, the $T_c(\varrho)$ minimum
corresponds to a
similar minimum around $P_0$ (lower panel of Fig.~\ref{fig:tc}).
In particular, the qualitative behaviour of $T_c(\varrho)$ and $T_c(P)$ is
identical, indicating that thermodynamic features governed by the equation
of state are not central to understanding the observed dynamics.

Since transport processes slow down dramatically upon approaching $T_c$,
relaxation in the silica melt along an isotherm first becomes \emph{faster}
with increasing pressure as the distance $|T-T_c|$ increases. Only once
the pressure exceeds $P_0$, this distance decreases and hence relaxation
becomes slower again.
For comparison, in the Lennard-Jones system, $T_c(\varrho)$ increases
monotonically with increasing density \cite{tvlj} which leads to the
expected monotonous decrease of diffusivity with increasing pressure.

\begin{figure}
\begin{indented}\item[]
\includegraphics[width=\linewidth]{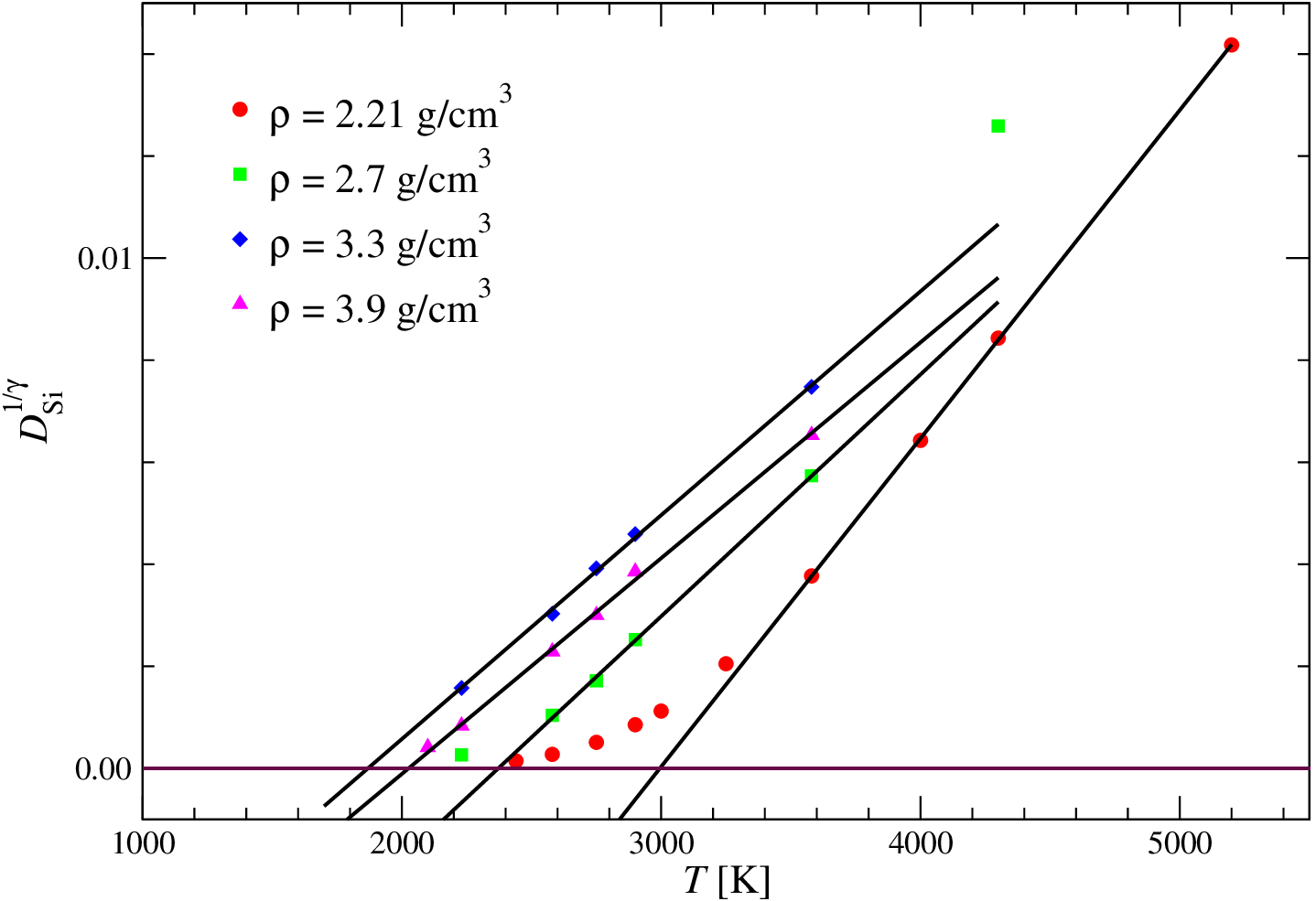}
\end{indented}
\caption{\label{fig:tcrect}%
  Rectification plot for the determination of $T_c(\varrho)$ from
  the MD-simulated $\text{Si}$ diffusion coefficients $D_{\text{Si}}$:
  $D_{\text{Si}}^{1/\gamma}$ with $\gamma=2.15$ for various densities as
  indicated (symbols). Solid lines are linear fits determining $T_c$
  as the intersection with the $T$ axis.
}
\end{figure}

To demonstrate the level of agreement between MCT and simulation data,
we have also estimated $T_c$ from the MD dynamics by fitting asymptotic power
laws of MCT to the temperature dependence of the diffusion coefficients.
The theory predicts $D\sim|T-T_c|^{\gamma}$, and while in principle
$\gamma$ will depend on density (see below), we anticipate that this change
will be relatively small and use $\gamma=2.15$ for all data sets.
Figure~\ref{fig:tcrect} shows exemplary rectification plots from
which $T_c(\varrho)$ is determined by linear fits to $D^{1/\gamma}$ (symbols)
in a restricted temperature range not too close and not too far above $T_c$,
shown as solid lines.
Deviations from a linear slope at higher temperatures result from
preasymptotic corrections, while those at lower temperatures signify
non-MCT relaxation process (`hopping'). In agreement with the observation
made in Fig.~\ref{fig:diff}, the latter deviations appear more dominant at
lower densities, indicating that the MCT description will be better for
pressurized silica than under atmospheric conditions.

Although the $T_c$ determined from simulation, shown as circle
symbols in the upper panel of Fig.~\ref{fig:tc}, are
systematically lower than the $T_c$ calculated within the theory, the
position of the minimum at $\varrho_0$ is in good agreement.
There is a slight tendency for the MCT result to better agree with
the MD-determined $T_c$ at higher densities, in line with the expectation
that MCT deals quantitatively better with dense liquids.
$T_c$ at ambient pressure has previously been determined using the BKS
potential of silica. MD simulations yielded $T_c=3330\,\text{K}$, while
MCT calculations without triplet correlation contributions gave
$T_c=3962\,\text{K}$ \cite{sciortino01}.
These values are significantly higher than our results, although the
static structure obtained from the BKS respectively the CHIK potential
shows relatively small differences \cite{Horbach.thisissue}. This
underlines the
importance of obtaining static-structure factor input for MCT as accurately as
possible.
The inclusion of the static triplet correlation function $c^{(3)}$
for the BKS potential led to a
better agreement in the length-scale dependence of the glass form factors
\cite{sciortino01}, but worsened the agreement for $T_c$. Whether the same
will also hold for the CHIK potential, is unclear, but a noticable shift
of the $T_c$ values presented in Fig.~\ref{fig:tc} has to be anticipated.

\begin{figure}
\begin{indented}\item[]
\includegraphics[width=\linewidth]{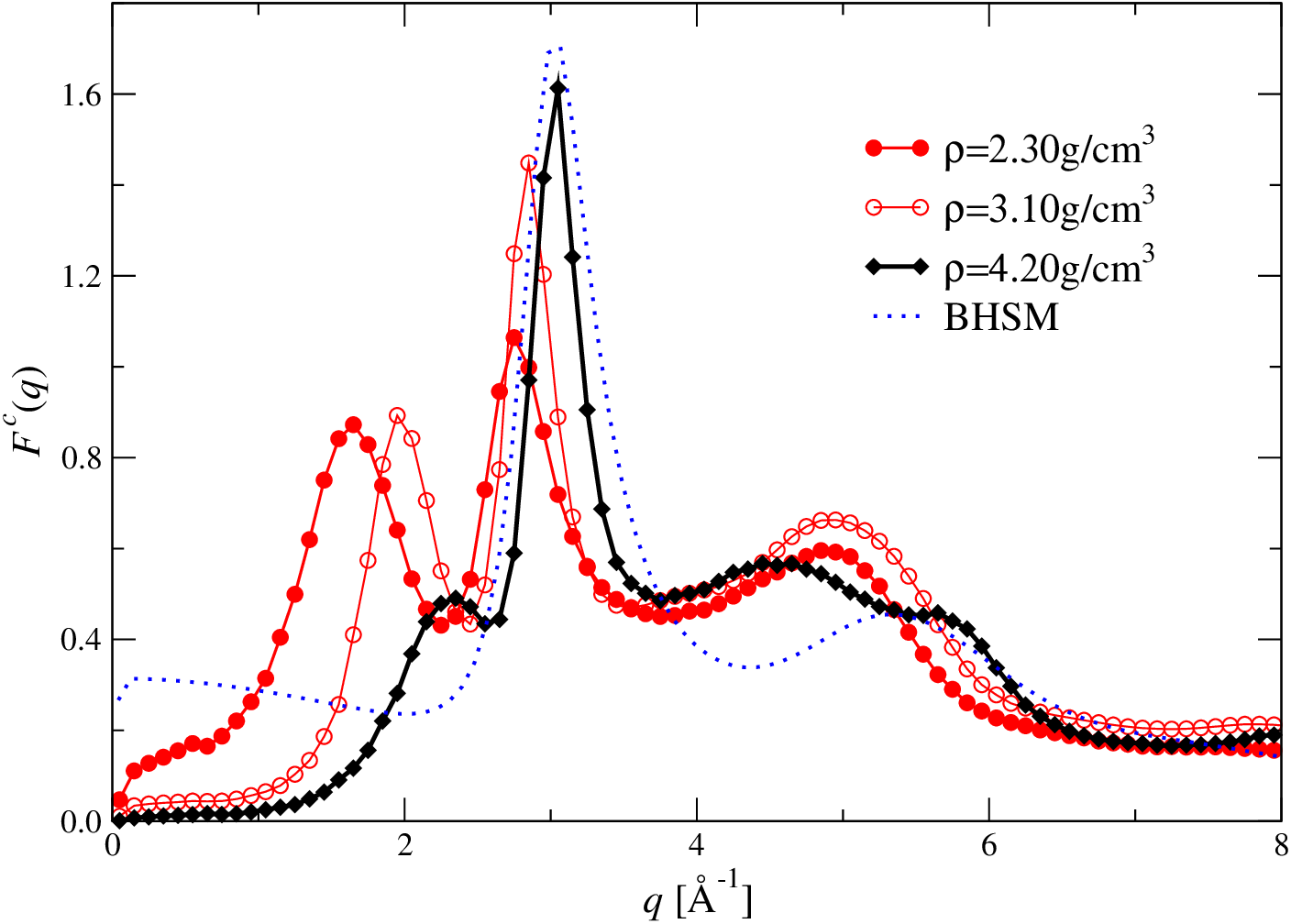}
\end{indented}
\caption{\label{fig:fq}%
  Nonergodicity parameters $F^c(q)$ calculated within MCT for model silica
  melts, at the critical temperature $T_c$ for densities
  $\varrho=2.3\,\text{g}/\text{cm}^3$ (solid circles),
  $3.1\,\text{g}/\text{cm}^3$ (open circles), and
  $4.2\,\text{g}/\text{cm}^3$ (diamonds). Shown for comparison is $F^c(q)$
  for a binary hard-sphere mixture (dotted line), see text for details.
}
\end{figure}

The variation in $T_c$ shown in Fig.~\ref{fig:tc} can be understood within
MCT by looking at specific features of the static structure factors.
While at the lower densities, all $S_{\alpha\beta}(q)$ show a pronounced
scattering peak at $q_1\approx1.7/\text{\AA}$ reflecting $\text{Si}\text{O}_2$
tetrahedra as the structural units of the system, this peak vanishes at
the expense of the main diffraction peak in $S_{\alpha\beta}(q)$, located at
$q_2\approx2.8/\text{\AA}$ at $\varrho=2.3\,\text{g}/\text{cm}^3$
and corresponding to typical interatomic, rather
than inter-tetrahedral distances \cite{Horbach.thisissue}.
This trend can be identified also in the
total density-density correlations, $S(q)=\sum_{\alpha\beta}x_\alpha x_\beta
S_{\alpha\beta}(q)$. Similarly, it is reflected in the glass form factors
$F(q)=\sum_{\alpha\beta}\lim_{t\to\infty}x_\alpha x_\beta
\Phi_{\alpha\beta}(q,t)$,
i.e., that part of density fluctuations which is frozen in upon crossing
the MCT glass transition. Figure~\ref{fig:fq} shows the critical form
factors $F^c(q)$ as a function of wave number, evaluated at several
densities along the MCT transition line $T_c(\varrho)$. While for
$\varrho=2.3\,\text{g}/\text{cm}^3$ (filled circles, corresponding to
$P\approx1.25\,\text{GPa}$), two peaks of almost identical height appear
at the two $q$-values $q_1$ and $q_2$, the first peak has almost disappeared
at $\varrho=4.2\,\text{g}/\text{cm}^3$ (diamond symbols, corresponding to
$P\approx46.2\,\text{GPa}$). At the same time, the peak at $q_2$ has grown.
This evolution reflects a gradual loss of tetrahedral ordering, and a
smooth crossover to a system showing signatures of a dense
liquid. Around the minimum in $T_c$, both contributions prevail, as the
$F^c(q)$ for $\varrho=3.1\,\text{g}/\text{cm}^3$ (corresponding to
$P\approx13.0\,\text{GPa}$) shown in Fig.~\ref{fig:fq} demonstrates.
Hence we attribute the initial decrease of $T_c(\varrho)$ to a loss of
chemical short-range order (the tetrahedral structure), and the subsequent
increase to an increase in nearest-neighbour cageing.
This crossover visible in $F^c(q)$ corresponds to a change
in mean coordination numbers for the $\text{Si}$ atoms in
$\text{Si}\text{O}_2$ with increasing pressure. MD simulations for example
show a gradual crossover from predominantly four-fold coordinated $\text{Si}$
atoms at $\varrho\approx2.3\,\text{g}/\text{cm}^3$ to a significant number of
five- and six-fold coordinated ones \cite{Horbach.thisissue}.

To demonstrate the approach to a frozen structure that resembles one
governed by packing effects, we show in addition in Fig.~\ref{fig:fq} the
$F^c(q)$ obtained from a binary hard-sphere mixture (using the Percus-Yevick
approximation for $\bs S(q)$) with diameters chosen as
$d_{\text{large}}=1.82\,\text{\AA}$ and $d_{\text{small}}=1.46\,\text{\AA}$
and concentration $x_{\text{large}}=1/3$. Considering that the
covalent radius of $\text{Si}$ yields
roughly $d_{\text{Si}}=2.2\,\text{\AA}$, the
value of $d_{\text{large}}$ appears reasonable if one takes into account
the complicated
interatomic potentials in the $\text{Si}\text{O}_2$ melt.
In particular, the comparison with the hard-sphere $F^c(q)$ shows that
besides the growing main peak, the emerging shoulder at
$q\approx5.7/\text{\AA}$ can be attributed to excluded-volume effects.
It can be anticipated that the $F^c(q)$ for the silica melt and the
hard-sphere mixture further approach each other, as the silica density is
further increased. Only in the $q\to0$ limit the two quantities show no
convergence, related to the fact that the two systems have rather different
isothermal compressibilities.

\begin{figure}
\begin{indented}\item[]
\includegraphics[width=\linewidth]{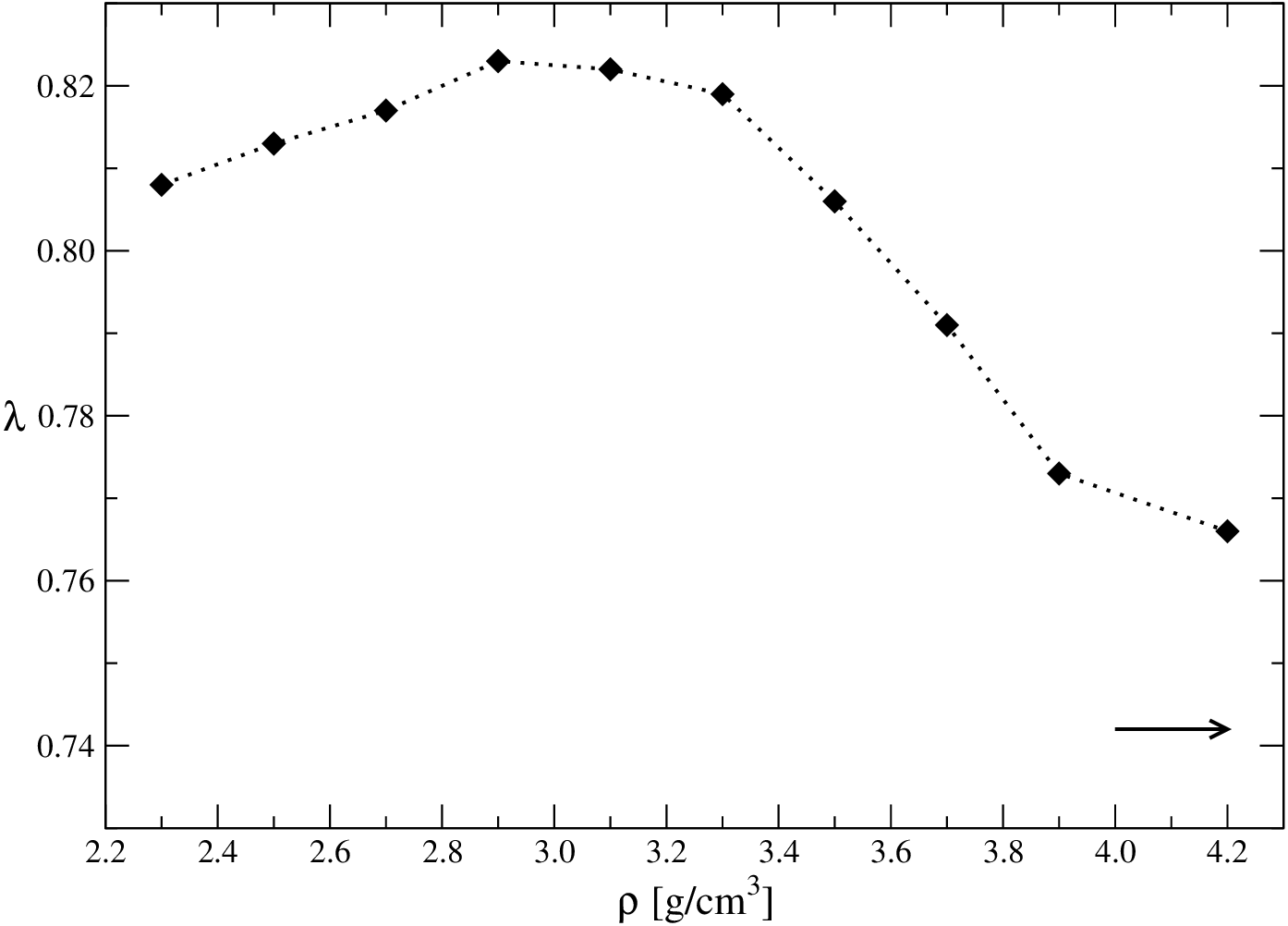}
\end{indented}
\caption{\label{fig:lambda}%
  MCT exponent parameter $\lambda$ calculated for silica as a function
  of density $\varrho$. The horizontal arrow indicates the value obtained
  for the binary hard-sphere mixture shown in Fig.~\protect\ref{fig:fq}.
}
\end{figure}

A further corroboration for the crossover from a tetrahedral network former
to an excluded-volume influenced glass comes from the analysis of the MCT
exponent parameter $\lambda$. This quantity controls the exponents of
the asymptotic expansions near the singulary $T_c$. It is bounded by
$1/2\le\lambda\le1$, but for common `fragile' glass formers, one usually
finds $\lambda\approx0.75$ as, e.g., for the Lennard-Jones model
\cite{goetze99}, and values $\lambda\ge0.8$ for systems where attractive
interactions play comparable role to excluded volume as, e.g., in the
square-well system \cite{Dawson2001}. The evolution
of $\lambda$ with increasing density in the present silica model
is shown in Fig.~\ref{fig:lambda}.
Interestingly, it shows values larger than $0.8$ for the lower densities
corresponding to moderate pressures. At high densities it systematically
decreases towards a value of $\lambda\approx0.766$
at $\varrho=4.2\,\text{g}/\text{cm}^3$. For comparison, the value
obtained for the binary hard-sphere mixture discussed above,
$\lambda\approx0.742$, is indicated in Fig.~\ref{fig:lambda} as a horizontal
arrow. This suggests that indeed the asymptotic dynamic behaviour of
the high-density silica melt slowly approaches that of a densely packed
mixture. Note also that maxima in $\lambda$, similar to the one shown
in the Fig.~\ref{fig:lambda}, have been argued to arise from an interplay
of two different arrest mechanisms \cite{Dawson2001}, consistent with
our picture of a gradual crossover with increasing pressure.
The parameter $\lambda$ in particular determines the exponent $\gamma$
for the asymptotic divergence of relaxation times or viscosities at $T_c$;
smaller values of $\lambda$ signify larger $\gamma$ and vice versa.

\begin{figure}
\begin{indented}\item[]
\includegraphics[width=\linewidth]{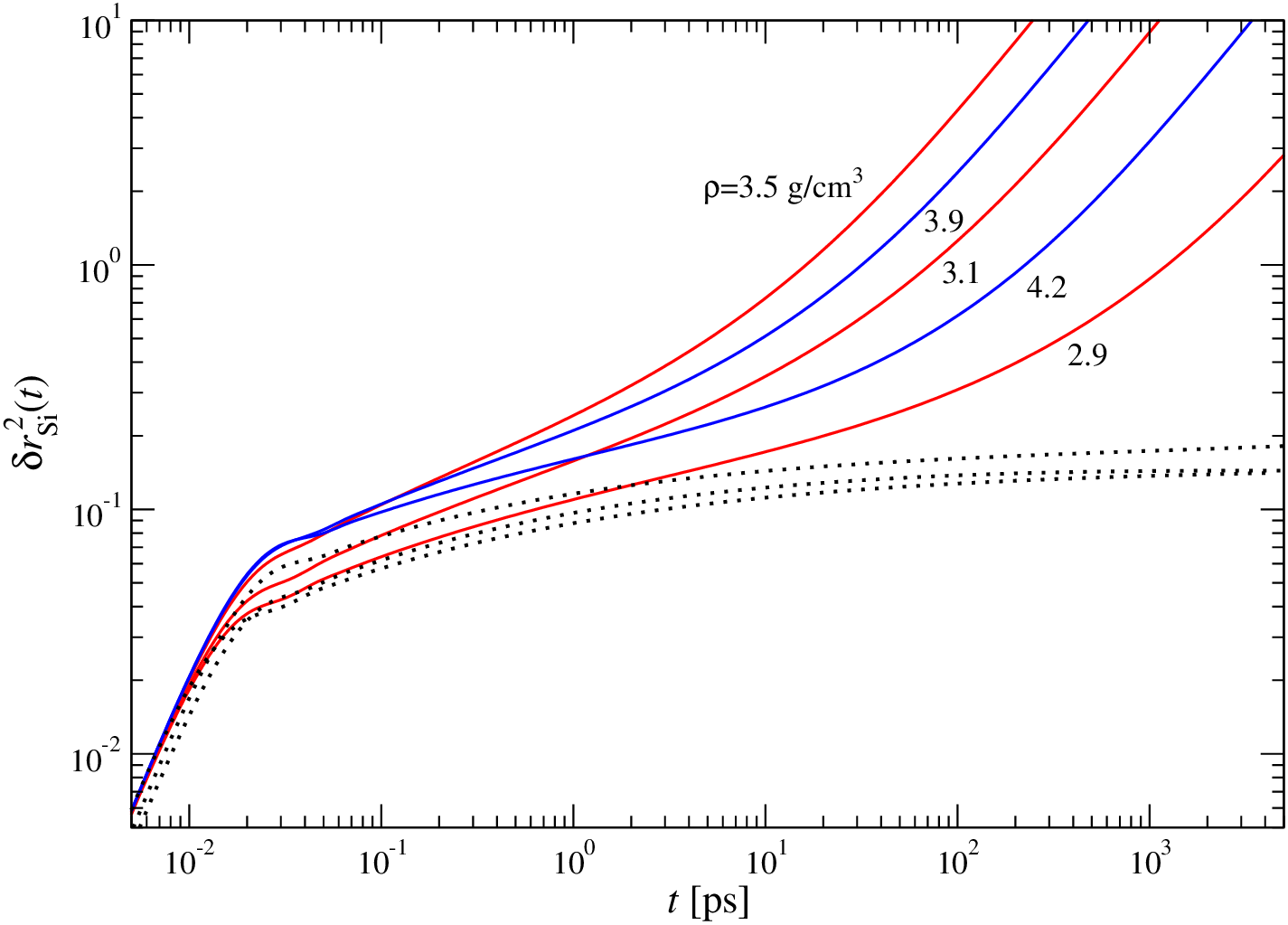}
\end{indented}
\caption{\label{fig:msd}%
  Mean-squared displacement of $\text{Si}$ atoms in the modeled silica melt,
  at various densities indicated and along the $T=2750\,\text{K}$ isotherm.
  Dotted lines indicate the corresponding quantities at $T=T_c(\varrho)$
  for $\varrho=2.7\,\text{g}/\text{cm}^3$, $3.5\,\text{g}/\text{cm}^3$, and
  $3.9\,\text{g}/\text{cm}^3$.
}
\end{figure}

The non-monotonous variation of $D_\alpha$ shown in Fig.~\ref{fig:diff}
leads to a corresponding non-monotonous variation in the mean-squared
displacement $\delta r^2_\alpha(t)$ (MSD) upon varying density or pressure
along an isotherm.  We show exemplary results for $\delta r^2_{\text{Si}}(t)$
at $T=2750\,\text{K}$ and various densities
in Fig.~\ref{fig:msd}. The variation at long times, i.e., for large
displacements $\delta r^2_{\text{Si}}\gtrsim1\,\text{\AA}$ reflects the
change in $D_\alpha$. But also at earlier times, in the $\text{ps}$ regime,
a strong non-monotonous variation in $\delta r^2_\alpha(t)$ remains.
Since asymptotically close to $T_c$, the plateau visible in the MSD is a
measure of the localization length of the individual particle, it is tempting
to read off from the intermediate-time window in Fig.~\ref{fig:msd} a
$\text{Si}$-localization length that shows an apparent change of almost
a factor $2$ as a function of density. This is what is also observed in
the MD simulation \cite{Horbach.thisissue}.
In fact, the value of, say, $\delta r^2_{\text{Si}}(1\text{ps})$
first increases with increasing density or pressure: $\text{Si}$ particles
are on that time scale less localized for higher pressure, reflecting
a change in the microscopic dynamics that becomes less dominated by
the strong localization within the tetrahedral network. However, this is
not the localization that is responsible for the MCT glass transition, and
which can only be read off from the MSD sufficiently close to $T_c$.
To this end, we show as dotted lines in Fig.~\ref{fig:msd} the
$\delta r^2_{\text{Si}}(t)$ at $T=T_c(\varrho)$ and observe that they agree
closely over the entire density range presented, indicating that the
cage localization length $r^c_{\text{Si}}\approx0.15\,\text{\AA}$ changes
only weakly as a function of pressure in this system.
If observed far from $T_c$, large preasymptotic corrections that are
particularly dominant for the MSD \cite{sperl}
give rise to the apparent shift in $r^c$.

For the $\text{O}$ atoms, the same qualitative trend as discussed in
connection with $\text{Si}$ holds, although the variation of $r^c_{\text{O}}$
is slightly larger. The ratio of localization lengths
$\delta=r^c_{\text{O}}/r^c_{\text{Si}}$ monotonically decreases from
$\delta\approx1.25$ at $\varrho=2.3\,\text{g}/\text{cm}^3$ to about
$\delta\approx1.05$ at $\varrho=4.2\,\text{g}/\text{cm}^3$, indicating
that the role played by the $\text{Si}$ and the $\text{O}$ atoms in the
dynamics assimilates at large pressure. Overall, both localization lengths
agree with Lindemann's criterion for melting,
stating that $r_\alpha^c\approx0.1d_\alpha$.

\section{Conclusions}

We have demonstrated that MCT, together with computer-simulation input
for the equilibrium liquid structure, reproduces a peculiar change
in the dynamics of a pressurized silica melt: upon increasing pressure,
atomic-scale transport as monitored through, e.g, self-diffusion coefficients,
first becomes faster. At a pressure around $20\,\text{GPa}$, a maximum
in diffusivity occurs, and at still higher pressures, transport starts
to slow down with increasing pressure. This is in broad agreement with
previous simulation data on model silica melts and with experiments on
various silicate mixtures.

MCT explains the diffusivity maximum in silica melts as arising from a
gradual change in the static structure on mesoscopic length scales,
where contributions connected to
tetrahedral ordering at wave number $q\approx1.7/\text{\AA}$ become
continuously less pronounced at the expense of contributions on the
length scale of the $\text{Si}$--$\text{Si}$ atom nearest-neighbour
distance, $q\approx 2.8/\text{\AA}$.
This is a feature found only in silica and similar network-forming melts,
whereas in simpler dense liquids such as the Lennard-Jones liquid, the
nearest-neighbour contribution in $S(q)$ remains dominant at all densities
or pressures. Hence, the latter `fragile' liquids do not show diffusivity
maxima. As an additional difference, the variation of $T_c(P)$ with
pressure in Lennard-Jones liquids is dominated by thermodynamic contributions
(viz., a strong variation in the equation of state) arising from the
presence of a gas--liquid spinodal \cite{tvlj}, while
in silica melts it is inherent to the slow glassy dynamics and not governed 
by the equation of state.

The qualitative correctness of MCT predictions for pressurized silica melts
is remarkable, since most experimental observations concern temperatures
well below the MCT $T_c$, where the theory in its present form is not
applicable. Nevertheless, the qualitative physical mechanisms responsible
for the anomalous pressure dependence of transport coefficients in this
strong glass former seem to be there already at far higher temperatures,
and captured in the MCT approximation.
A similar conclusion might apply to the distinction between `strong' and
`fragile' behaviour of the viscosity $\eta(T)$ around $T_g$. While MCT
cannot be applied there, it does yield for silica a gradual crossover of the
exponent $\gamma$ governing the initial increase in viscosity above $T_c$,
changing from a slower increase at ambient pressure to a stronger one
(more akin to a `fragile' liquid) at high pressures.

Based on this observation, we suggest that MCT can be used to investigate
in more detail the connection between `strong' glass formers with 
network-like structures at low pressures and `fragile' ones which are
characterized by a dense arrangement of constituent atoms.

\ack{
We thank for support by Schott Glas and for a generous grant of computing
time on the JUMP at the NIC J\"ulich.
}

\bibliography{lit}
\bibliographystyle{iopart-num}

\end{document}